\documentclass[prd,nofootinbib,floats,aps,twocolumn,tightenlines,superscriptaddress,floatfix]{revtex4-1}

\usepackage{graphicx}
\usepackage{graphics}
\usepackage[retainorgcmds]{IEEEtrantools}
\usepackage{color}
\usepackage{xcolor}
\usepackage{multirow}
\usepackage{xspace} 
\usepackage{slashed}
\usepackage[colorlinks=true, citecolor=blue,urlcolor=blue]{hyperref}
\usepackage{graphicx,marvosym,subfigure,amssymb,amsmath}
\usepackage{comment}
\usepackage{graphicx}
\usepackage[normalem]{ulem}
\usepackage[normalem]{ulem}
\usepackage{footmisc}

\newcommand{\coupl}{\zeta}

\newcommand{\mt}{\mathbb{M}_2}
\newcommand{\st}{{\mathbb{S}^2}}

\newcommand{\xp}{{x^\prime}}
\newcommand{\tp}{{t^\prime}}
\newcommand{\yp}{{y^\prime}}

\newcommand{\thp}{{\theta^\prime}}
\newcommand{\phip}{{\varphi^\prime}}

\newcommand{\be}{\begin{equation}}
\newcommand{\ee}{\end{equation}}
\newcommand{\pd}[2]{\frac{\partial#1}{\partial#2}}
\newcommand{\order}[1]{\mathcal{O}\left(#1\right)}
\newcommand{\evalO}[1]{\left(#1\right)_O}

\begin{document}
\global\parskip 6pt

\author{David Q. Aruquipa}
\email{daruquipa@cbpf.br}
\affiliation{Centro Brasileiro de Pesquisas F\'isicas (CBPF),  Rio de Janeiro, 
CEP 22290-180, 
Brazil}

\author{Marc Casals}
\email{marc.casals@uni-leipzig.de,mcasals@cbpf.br, marc.casals@ucd.ie.}
\affiliation{Institut f\"ur Theoretische Physik, Universit\"at Leipzig,  Br\"uderstra\ss e  16, 04103 Leipzig, Germany}
\affiliation{Centro Brasileiro de Pesquisas F\'isicas (CBPF),  Rio de Janeiro, 
CEP 22290-180, 
Brazil}
\affiliation{School of Mathematics and Statistics,
University College Dublin, Belfield, Dublin 4, D04 V1W8, Ireland}

\title{Hadamard Tail from 
Initial Data on the Light Cone}

\begin{abstract} 
Field perturbations of a curved background spacetime generally propagate not only at the speed of light but also at all smaller velocities. This so-called \textit{Hadamard tail} contribution to wave propagation is relevant in various settings, from classical self-force calculations to communication between quantum particle detectors. One method for calculating this tail contribution is by integrating the homogeneous wave equation using Characteristic Initial Data on the light cone. However, to the best of our knowledge, this method has never been implemented before except in the special case of conformally-flat spacetimes, where null geodesics emanating from a point do not cross. In this work, we implement this method on Pleba{\'n}ski-Hacyan spacetime, $\mt\times\st$, a black hole toy model which has caustics. We obtain new results in this spacetime by calculating the Hadamard tail of a scalar field everywhere where it is defined (namely, in the maximal normal neighbourhood of an arbitrary point) and investigate how it varies for various values of the coupling constant. This serves as a  proof-of-concept for the Characteristic Initial Data method on spacetimes where null geodesics emanating from a point {\it do} cross.
\end{abstract}

\date{\today}
\maketitle



\section{Introduction}

Field perturbations of a curved background spacetime obey a wave equation which dictates that they propagate ``mainly" along null geodesics. However, in general, there is also a part of the field, called the {\it tail}, which propagates slower than light (even in the case of a massless field). In other words, the  strong Huygens principle is generally violated in curved spacetimes~\cite{McLenaghan:1974,CzaporMcLenaghan}. This tail  is important for different reasons. For example, it is useful for calculating the self-force~\cite{Poisson:2011nh} on a point particle via the method of matched expansions~\cite{Anderson:Wiseman:2005}, where the retarded Green function of the wave equation needs to be regularized~\cite{CDOW13,PhysRevD.89.084021}. Also, the tail may give rise to an interesting relevant contribution  to the communication between quantum particle detectors~\cite{blasco2015violation,Jonsson:2020npo}.

Mathematically, the tail term may be defined in local neighbourhoods of spacetime points via the Hadamard form~\cite{Hadamard} of the retarded Green function. Specifically,  the tail is the term in the Hadamard form which has support only {\it inside} the (past) light cone of the field point. Where the tail term is nonzero, and henceforth focusing on the case of a scalar field for simplicity, it is equal to a bitensor $V(x,x')$. This tail bitensor satisfies the {\it homogeneous} wave equation, constrained  by its value on the light cone, which satisfies a transport equation (along null geodesics) with a given initial condition~\cite{Poisson:2011nh}.

The Hadamard tail bitensor $V(x,x')$ has been obtained in closed form in only a very few settings of high symmetry - specifically, and to the best of our knowledge, only when the background spacetime is flat~\cite{M&F} or conformally-flat\footnote{In these cases, the field considered is non-conformal, so that there exists a nontrivial tail.}, such as simple (spatially-flat) cosmological model spacetimes~\cite{Burko:2002ge,haas2005mass}, including (a patch of) de Sitter~\cite{Friedlander}. A simplifying feature of flat and conformally-flat spacetimes is that null geodesics emanating from a point do not cross each other. Furthermore, the maximal symmetry of, in particular, both flat  and de Sitter spacetimes means that it is possible to rewrite the  partial differential wave equations in these spacetimes as {\it ordinary} differential equations (where the independent variable is the geodesic distance), which are much easier to solve. However, in other spacetimes which are less symmetric  and where null geodesics emanating from a point do cross, such as black hole spacetimes, no closed form expression for $V(x,x')$ is known and, instead, one needs to resort to numerical or approximating analytical techniques. Focusing on black hole spacetimes, some of these techniques have been quite successful in calculating $V(x,x')$ in Schwarzschild spacetime~\cite{CDOWb,Ottewill:2009uj,Decanini:Folacci:2005a,CDOW13,PhysRevD.89.084021} but face much more significant difficulties in the case of Kerr. It is thus important to develop alternative methods for calculating the tail term.

In this paper we pursue the endeavour of calculating $V(x,x')$ by directly (numerically) integrating the (homogeneous) wave equation with given Characteristic Initial Data (CID) on the light cone. Specifically, we apply this method to the case of a scalar field propagating on Pleba{\'n}ski-Hacyan spacetime  (PH), $\mt\times\st$~\cite{Griffiths&Podolsky}. From a physical point of view, this spacetime serves as a black hole toy-model and it  captures the important feature that null geodesics emanating from a point do cross (in fact, similarly to Schwarzschild, there exist caustics where an $\st$-envelope of null geodesics focus). In its turn, from a technical point of view, the fact that PH is not a maximally-symmetric spacetime but is the direct product of two (two-dimensional) maximally-symmetric spacetimes (namely, $\mt$ and $\st$) means that its wave equation, while it is not an {\it ordinary} differential equation as in flat or de Sitter spacetimes, is reduced to a {\it two}-dimensional PDE; furthermore, the value of $V(x,x')$ on the light cone is known in closed form~\cite{Casals:2012px}. Our calculation of $V(x,x')$ in the specific case of a massles field with a coupling constant value of $\xi=1/8$ and  $x$ and $x'$ on static paths, agrees with~\cite{Casals:2012px}, where $V$ is calculated via a completely different method which involves infinite sums and integrals. This provides a check of our method and calculation and serves as a proof-of-concept for this method. We also obtain new results for $V$ in PH: for {\it any} pair of spacetime points where it is defined for $\xi=1/8$ as well as for $\xi=0$, $1/6$, $1/4$ and $1/2$.

For solving the homogeneous (two-dimensional) wave equation in PH by evolving CID we use a finite-difference scheme.  Refs.~\cite{mark2017recipe,Lousto:1997wf} proposed and implemented a CID scheme for obtaining the multipolar {\it modes} of the field (or retarded Green function) in Schwarzschild spacetime. We adapted this scheme to calculate the full $V(x,x')$ in PH spacetime. We note that some peculiarities of the PDE satisfied by $V$ in PH will also be present in the PDE satisfied by $V$ in Schwarzschild, so that our adaptation of the scheme will probably be useful for any future investigation in the latter spacetime. Furthermore, we developed the scheme to higher order than in~\cite{mark2017recipe,Lousto:1997wf}.
Throughout the paper we adopt the   convention that, unless otherwise explicitly specified, the ``order" of a scheme will refer to the {\it local} truncation error (LTE); thus, in particular, a ``third (fourth) order scheme'' will refer to a numerical scheme with LTE of order three (four). 

The rest of this paper is organized as follows. In Sec.~\ref{sec:Hadamard} we introduce the Hadamard form and give the explicit forms of the scalar wave equation and some Hadamard quantities  in PH. In Sec.~\ref{sec:CID} we present the characteristifc initial value problem in PH, our finite difference (third and fourth order) schemes for solving it and the results of our calculations. 
We conclude the main part of the text in Sec.~\ref{sec:Discussion} with a brief discussion. Finally, we have two appendixes. In App.~\ref{app:taylorCoeffs} we provide extra equations which were needed for deriving a scheme in the main text but not for implementing it. In App.~\ref{app:nextOrderScheme} we lay out the ground work for developing a scheme of order higher (fifth, and even, sixth) than in the main text.

We choose units such that $G=c=1$.


\section{Wave equation, Hadamard Form and Tail}\label{sec:Hadamard}


\subsection{A general spacetime}

A scalar field perturbation of a background spacetime satisfies a wave equation. Specifically, its retarded Green function satisfies:
\be\label{eq:GF eq}
\left(\Box-m^2-\xi R\right)G_{\textrm{ret}}(x,\xp)=-4\pi\delta_4(x,\xp),
\ee
where $m$ is the mass of the field, $R$ is the Ricci scalar, $\xi$ is a coupling constant and $x$ and $\xp$ are, respectively, the field  and  base spacetime points.

The Hadamard form provides an analytic expression for the singularities of the retarded Green function when $x$ is in a local (normal\footnote{A normal neighbourhood of $\xp$ is a region containing $\xp$ such that every $x$ in that region is connected to $\xp$ by a unique geodesic which lies within the region.}) neighbourhood of  $\xp$~\cite{DeWitt:1960,Friedlander,Poisson:2011nh}:
\begin{align} & G_{\textrm{ret}}(x,\xp)=
\label{grhad} \\ & [U(x,\xp)\delta(\sigma(x,\xp))+V(x,\xp)\theta(-\sigma(x,\xp))]\theta_+(x,\xp),
\nonumber
\end{align}
where $\delta$ and $\theta$ are, respectively, the  Dirac delta and Heaviside distributions, \[\theta_+(x,\xp)\equiv \left\{\begin{array}{ll} 1 & \hbox{if $x$ lies to the future of $\xp$},\\ 0 & \hbox{otherwise},\end{array}\right. \] and $U$ and $V$ are biscalars which are smooth in that local neighbourhood. Here, $\sigma(x,\xp)$ is Synge’s world-function, i.e., one-half of the square of the geodesic distance along the unique geodesic connecting $\xp$  and $x$. Thus, clearly, the term with $U$ in Eq.~\eqref{grhad} has support only {\it on} the light cone whereas the term with $V$ has support {\it inside} the light cone: this is the tail term which is the focus of this paper.

The Hadamard tail biscalar $V$ satisfies the {\it homogeneous} wave equation:
\be\label{eq:V wave eq}
\left(\Box-m^2-\xi R\right)V(x,\xp)=0,
\ee
constrained by its value on the light cone:
\be\label{eq:transp eq V}
\hat{V}_{,\alpha}\sigma^{\alpha}+\frac{1}{2}\left(\sigma^{\alpha}{}_{\alpha}-2\right)\hat{V}=\frac{1}{2}\left(\Box-m^2-\xi R\right)\left.U\right|_{\sigma=0},
\ee
where $\hat{V} \equiv V|_{\sigma=0}$ and  $\sigma^{\alpha}{}_{\alpha}\equiv\nabla_\alpha\nabla^\alpha\sigma$. Eq.\eqref{eq:transp eq V} is in fact a transport equation along a light cone-generating null geodesic. It is to be solved together with the initial condition corresponding to the value of $V$ at coincidence (i.e., at $\xp=x$):
\be\label{eq:IC V}
V(x,x)=\frac{1}{12}\left(1-6\xi\right)R(x)-\frac{1}{2}m^2,
\ee
which follows partly from the fact that  $V$ is smooth at coincidence. Thus, the tail biscalar $V(x,x')$ satisfies the initial value problem consisting of the  PDE \eqref{eq:V wave eq} together with the CID provided by solving the transport equation \eqref{eq:transp eq V} together with the initial condition \eqref{eq:IC V}. This guarantees that the solution $V(x,x')$ exists and is unique.

One method for trying to calculate $V(x,\xp)$ is to  express it as an asymptotic series:
\be 
V(x,x')=\sum_{n=0}^\infty \nu_n(x,x')\sigma^n\label{eqn:VsigmaExpansion},
\ee
where the coefficients $\nu_n(x,x')$ satisfy certain recurrence relations in the form of transport equations~\cite{DeWitt:1960,Decanini:Folacci:2005a}. Ref.~\cite{Ottewill:2009uj} provided a complete procedure for calculating $\nu_n$ to arbitrary $n$ by solving a system of transport equations. Unfortunately, however, as the coefficient order $n$ increases, these transport equations become increasingly hard to solve (even numerically and for low $n$). Furthermore, although the series in Eq.~\eqref{eqn:VsigmaExpansion} converges uniformly in subregions of normal neighbourhoods~\cite{DeWitt:1960,Friedlander}, it is not actually guaranteed to converge in the whole maximal normal neighbourhood of a point. A more practical method for calculating $V(x,x')$ in spherically-symmetric spacetimes is to expand this bitensor in small {\it coordinate} distance between $x$ and $x'$~\cite{CDOWb}. Although this method has proven to be very useful in Schwarzschild spacetime~\cite{CDOW13,PhysRevD.89.084021}, it is naturally adapted to spherical symmetry and so it still needs to be developed in Kerr spacetime.

In this paper, we shall instead return to the original initial value problem for calculating the Hadamard tail. Specifically, known Hadamard coefficients\footnote{These coefficients are not independent from each other, as we later explain.} $\nu_n$ in the series in Eq.~\eqref{eqn:VsigmaExpansion} shall provide  the CID which we shall then use in a numerical scheme for solving the full wave Eq.~\eqref{eq:V wave eq} for $V(x,x')$ for {\it any} pair of points in PH spacetime.


\subsection{PH spacetime}\label{sec:PH}
PH spacetime is the direct product of two-dimensional Minkowski spacetime $\mt$ and the two-sphere $\st$~\cite{Griffiths&Podolsky}. This becomes manifest when writing its line element as\footnote{We make the units choice that the radius of the two-spheres is equal to one.}
\be ds^2=-dt^2+dy^2+d\Omega^2,\label{PHlel}\ee where $$d\Omega^2=d\theta^2+\sin^2\theta d\varphi^2,$$ with $(t,y)\in \mathbb{R}^2$ global inertial coordinates in  $\mt$, and $\theta\in [0,\pi]$ and $\varphi\in (-\pi,\pi]$ the standard angular coordinates in $\st$. The Ricci scalar of PH is $R=2$.

PH being the direct product $\mt\times \st$, its  world function $\sigma$ is readily given~\cite{Casals:2012px} as the sum of the world functions in $\mt$ and  $\st$, respectively $\sigma_{\mt}$  and $\sigma_{\st}$: $\sigma(x,\xp) = \sigma_{\mt}+\sigma_{\st}$. In their turn, these world functions are given, in normal neighbourhoods, by
\be \sigma_{\mt}=-\frac12\eta^2\equiv -\frac12(t-\tp)^2+\frac12(y-\yp)^2 \label{sigbardef}\ee
and
\be 
\sigma_{\st}=\frac{\gamma^2}{2},
\ee
where
\be
\cos\gamma\equiv \cos\theta\cos\thp+\sin\theta\sin\thp\cos(\varphi-\phip).\label{gamdef}
\ee
We thus have
\be 
\sigma=
-\frac12\eta^2 + \frac12\gamma^2=-\frac12(t-\tp)^2+\frac12(y-\yp)^2+ \frac12\gamma^2.
\ee
Clearly, $\eta\in \mathbb{R}$ is the geodesic distance in the whole of $\mt$. In its turn,  $\gamma\in [0,\pi]$ is the geodesic (or angle) {\it separation} in  $\st$, while it also is the geodesic distance in normal neighbourhoods of $\st$ (see~\cite{Casals:2019heg,casals2016global} for this subtle but important distinction between geodesic separation and geodesic distance in the context of Schwarzschild spacetime). Null geodesics (for which $\sigma=0$ in normal neighbourhoods) focus at the first caustic points: $\eta=\gamma=\pi$. After crossing the first caustic, the envelope of null geodesics emanating from a base point  $\xp$ forms the (future) boundary of the maximal normal neighbourhood of the base point; this boundary is given by $\eta=2\pi-\gamma\in [\pi,2\pi]$  (see the left panel of Fig.1 in \cite{Casals:2012px}). Since $V(x,x')$ is only defined in normal neighbourhoods, this will also be part of the boundary of the grid in our numerical scheme. The other part  is given by the null hypersurface corresponding to the envelope of future-directed {\it direct} null geodesics, i.e., by $\eta=\gamma\in [0,\pi)$ (so that it is $\sigma=0$ with $\eta\geq 0$). That is, the future boundary $\eta=2\pi-\gamma\in [\pi,2\pi]$  of the maximal normal neighourhood of an arbitrary base point together with the boundary $\eta=\gamma\in [0,\pi)$ of the causal future of the base point form the boundary of our numerical grid.

It is straight-forward to obtain the d'Alembertian in PH in the above coordinates:
\begin{align}\label{eqn:BoxM2S2}
&
\Box=\Box_{\mt}+\Box_{\st},\\
&
\Box_{\mt}=-\frac{\partial^2}{\partial t^2}+\frac{\partial^2}{\partial y^2},
\nonumber\\
&
\Box_{\st}=\frac{1}{\sin\theta}\frac{\partial}{\partial \theta}\sin\theta\frac{\partial}{\partial \theta}+\frac{1}{\sin^2\theta}\frac{\partial^2}{\partial\varphi^2}.\nonumber
\end{align}

Now, given that both  $\mt$ and  $\st$ are maximally-symmetric manifolds, it is easy to see that the above operators $\Box_{\mathbb{M}_2}$ and $\Box_{\mathbb{S}^2}$ become {\it ordinary} differential operators when rewritten in terms of the corresponding geodesic distances. Explicitly,
\be
\Box_{\mathbb{M}_2}=-\frac{\partial^2}{\partial \eta^2}-\frac{1}{\eta}\frac{\partial}{\partial \eta}
\ee
and
\begin{align}
\Box_{\mathbb{S}^2}=\frac{\partial^2}{\partial \gamma^2}+\cot\gamma \frac{\partial}{\partial \gamma}=
\frac{1}{\sin\gamma}\frac{\partial}{\partial \gamma}\left(\sin\gamma\frac{\partial}{\partial \gamma}\right).
\end{align}

The wave equation~\eqref{eq:V wave eq} thus becomes
\be\label{eq:wave eq PH}
\left[
\frac{\partial^2}{\partial \gamma^2}+\cot\gamma \frac{\partial}{\partial \gamma}
-\frac{\partial^2}{\partial \eta^2}-\frac{1}{\eta}\frac{\partial}{\partial \eta}
-\coupl\right]\! \! V(x,\xp)\!=0,
\ee
where $\coupl\equiv m^2+\xi R=m^2+2\xi$. Thus,  in PH, we have reduced the  wave equation, which is generally a four-dimensional PDE, to a {\it two}-dimensional PDE.

In~\cite{Casals:2012px}, it was found that, in PH, it is $V=V(\eta,\gamma)$ and $\nu_n=\nu_n(\gamma)$, $\forall n\ge 0$, \footnote{Even though we wrote $\nu_n=\nu_n(x,x')$ in Eq.~\eqref{eqn:VsigmaExpansion}, the obvious change of notation in its arguments to $\nu_n=\nu_n(\gamma)$  is justified by the symmetries of PH.} and closed form expressions for some Hadamard quantities were obtained. Specifically, it was found that
\begin{align} \label{udef}
&
U(x,x')= U(\gamma)=
\left|\frac{\gamma}{\sin\gamma}\right|^{1/2},
\end{align}
and, by solving Eqs.~\eqref{eq:transp eq V} and \eqref{eq:IC V}, that
\begin{align}
\hat{V} =\nu_0(\gamma)=\frac18U(\gamma)\left(1-4\coupl
+\frac{1}{\gamma^2}-\frac{\cot\gamma}{\gamma}\right).\label{v0def}
\end{align}
The higher orders $\nu_n$, $n>0$, can in principle be obtained from $\nu_{n-1}$ via a recurrence relation. For the specific case $\coupl=1/4$ it was obtained that
\begin{align} \label{eq:hat V1 M2xS2}
& \nu_1=
\\ &
U(\gamma)\frac{2\gamma^2-3 \csc ^2(\gamma ) \left[6 \gamma ^2+2 \gamma \sin (2 \gamma )+5 \cos (2 \gamma )-5\right]}{256 \gamma ^4},
\nonumber
\end{align}
for which regularity of $\nu_1$ at $\gamma=0$ was required.

We note that $\hat{V}$ is regular for all $\gamma\in [0,\pi)$ but it diverges (like $(\pi-\gamma)^{-3/2}$) at the antipodal points $\gamma=\pi$. These antipodal points, however, lie outside maximal normal neighbourhoods: As is manifest, $\st$-envelopes of null geodesics focus along $\gamma=\pi$ (a line of caustics), as in Schwarzschild spacetime. It has been observed~\cite{Dolan:2011fh,Zenginoglu:2012xe,casals2016global,Ori1,Harte:2012uw,Casals:2012px,CDOWa} that, when this happens, the retarded Green function diverges when $\xp$ and $x$  are connected by a null geodesic (even beyond normal neighbourhoods) displaying the following {\it global} fourfold (leading) singularity structure\footnote{Here, $\sigma$ refers to a well-defined extension of the world function outside normal neighbourhoods~\cite{casals2016global,Casals:2019heg}. We also note that this structure does not hold at caustics~\cite{casals2016global} and that the subleading order (in Schwarzschild and outside caustics) is given in~\cite{casals2016global}.}: $\delta(\sigma) \to \text{PV}\left(1/\sigma\right)\to -\delta(\sigma) \to -\text{PV}\left(1/\sigma\right)\to \delta(\sigma)\dots$, where $\text{PV}$ denotes the principal value distribution. Since $V$ is equal to $G_{\textrm{ret}}$ in a region of causal separation which lies inside a normal neighbourhood, one expects $V$ to diverge like $G_{\textrm{ret}}$, i.e. as $\text{PV}\left(1/\sigma\right)$, when approaching the end of the normal neighbourhood in this direction. As we shall see in Sec.~\ref{sec:results V}, the divergence of $\hat{V}$ at $\gamma=\pi$ (see Eq.~\eqref{v0def}) propagates in this manner throughout the end of the maximal normal neighbourhood of the base point.

The upshot is that, in PH, we have reduced the original wave equation  to a two-dimensional PDE (see Eq.~\eqref{eq:wave eq PH}), and that the Hadamard tail $\hat{V}$ on the light cone is known in closed form  (see Eq.~\eqref{v0def}, as well as the higher order in Eq.~\eqref{eq:hat V1 M2xS2}). In the next section we will use these advantageous features to numerically solve the wave equation \eqref{eq:wave eq PH} and thus to calculate $V$ inside the light cone.


\section{Solving the Characteristic Initial Value Problem for the Hadamard Tail}\label{sec:CID}

In this work, we will directly solve the wave equation \eqref{eq:wave eq PH} as a characteristic initial value problem for the tail biscalar $V$. More concretely, we shall develop a numerical scheme which will evolve initial data on the light cone, i.e., on
$$
    \sigma=-\frac{1}{2}\eta^2+\frac{1}{2}\gamma^2=0.
$$
This CID is given by Eqs.\eqref{eqn:VsigmaExpansion}, \eqref{v0def} and \eqref{eq:hat V1 M2xS2}. This section provides the details of the  scheme and the results. We split this section into three subsections: we first rewrite the wave equation in variables suitable to the CID problem; we then describe the numerical scheme; finally, we show our results for $V$.

\subsection{Wave equation as a Characteristic Initial Value problem} 

Let us  introduce the variables 
\be\label{eq:uv}
u\equiv \eta-\gamma, \quad v\equiv \eta+\gamma,
\ee
which are naturally adapted to the characteristic initial value problem, since $\sigma=-uv/2$. In these variables, the d'Alembertian in PH (see Eq.~\eqref{eq:wave eq PH}) becomes
\be
    \Box=-4\frac{\partial^2}{\partial u\partial v}-Q\frac{\partial}{\partial v}-S\frac{\partial}{\partial u},
\ee
where
\begin{align*}
    Q\equiv \,&\frac{2}{v+u}-\cot\frac{v-u}{2},\\
    S\equiv\,&\frac{2}{v+u}+\cot\frac{v-u}{2}.
\end{align*}

Therefore, Eq.~\eqref{eq:wave eq PH} turns into
\be
    \left(4\frac{\partial^2}{\partial u\partial v}+Q\frac{\partial}{\partial v}+S\frac{\partial}{\partial u}+\coupl\right)V(x,\xp)=0.\label{eqn:VUVEqn}
\ee
We note the appearance of {\it first}-order derivatives with respect to $u$ and $v$, arising from the first-order derivatives with respect to $\eta$ and $\gamma$ in Eq.~\eqref{eq:wave eq PH}. We also note that, even though $u$ and $v$ range over the reals to cover the whole spacetime, the domain over which we solve Eq.~\eqref{eqn:VUVEqn} is limited by the range of $\gamma\in [0,\pi]$ and the region where $V$ is defined. In the next subsection we detail how these constraints are reflected on the domains for $u$ and $v$.

Let us now turn to the CID. On the $u-v$ plane (see Fig.~\ref{fig:CIDGridM2xS2}), the light cone of the origin $=u=v=0$ (i.e., $\eta=\gamma=0$, which corresponds to coincidence, $x=\xp$) is located along the $u=0$ and $v=0$ lines. Thus, $\hat{V}$ in terms of $u$ and $v$, is given by
\begin{equation}\label{eqn:VCIDBC}
    \begin{aligned}
    \left.V\right|_{u=0}=\,&\nu_0\left(\frac{v}{2}\right),\\
    \left.V\right|_{v=0}=\,&\nu_0\left(-\frac{u}{2}\right),
    \end{aligned}
\end{equation}
where $\nu_0=\nu_0(\gamma)$ is given in Eq.~\eqref{v0def}.

The wave Eq.~\eqref{eqn:VUVEqn}, together with the CID \eqref{eqn:VCIDBC}  constitutes our characteristic initial value problem. We next present the finite difference method that we shall use to   solve it.

\subsection{Numerical Scheme}

\begin{figure}
    \centering
    \includegraphics{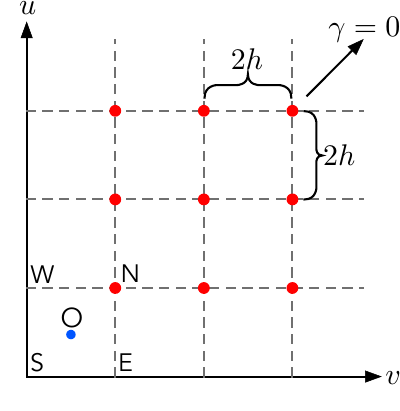}
    \caption{
    Grid distribution for a finite difference scheme for solving a two-dimensional PDE where $u$ and $v$ denote the independent variables and $2h$ is the stepsize.
    }
    \label{fig:CIDGridM2xS2}
\end{figure}

There have already been some implementations of CID schemes for solving the two-dimensional PDE (not containing first order derivatives, unlike Eq.~\eqref{eqn:VUVEqn}) which is obeyed by the (smooth factor\footnote{The other factor contains  non-smooth Heaviside distributions.} in the) $\ell$-multipolar {\it modes} of the retarded Green function in Schwarzschild spacetime (see, e.g., Eq.~(C2) in~\cite{Jonsson:2020npo}). Let us briefly discuss these schemes. In Ref.~\cite{mark2017recipe}, the authors implemented a fourth order scheme (previously proposed by Lousto and Price \cite{Lousto:1997wf}), that is, with LTE of order $h^4$, where $2h$ is the stepsize of the grid (see Fig.~\ref{fig:CIDGridM2xS2}). In order to calculate the value of a field mode  at a point, their scheme required field mode data on the immediately ``previous" grid points -- e.g., on the points S, E and W in order to obtain the value of the field mode at the point $N$ in Fig.~\ref{fig:CIDGridM2xS2}.
In~\cite{Jonsson:2020npo}, together with collaborators, we  extended the scheme in~\cite{mark2017recipe} from fourth to sixth order at the expense of requiring the value of the field mode  and of its first-order derivatives at the same grid points as in the lower-order scheme of~\cite{mark2017recipe}.
More recently, Ref.~\cite{PhysRevD.103.124022} came up with another scheme which they implemented to sixth order (although in principle can be generalized to any higher order) at the expense of requiring the value of the field mode at more points than in Refs.~\cite{mark2017recipe,Jonsson:2020npo}.

In our current work in PH, the PDE \eqref{eqn:VUVEqn} is also two-dimensional but, unlike in the case of Schwarzschild just reviewed,  and as we have emphasized, it contains first order derivatives and it is satisfied by the {\it full} field (instead of multipolar modes). We note that first order derivatives also appear in Schwarzschild spacetime  in the PDE satisfied by the full field as well as in the Teukolsky PDE satisfied by multipolar field modes for fields of nonzero spin~\cite{Teukolsky:1973ha}. For solving  Eq.~\eqref{eqn:VUVEqn}, we choose to essentially follow the  fourth order scheme of Ref.~\cite{Jonsson:2020npo} and adapt it to our specific PDE for the full field instead of just its modes.

Another difference between our setup and that in Refs.~\cite{mark2017recipe,Jonsson:2020npo,PhysRevD.103.124022} is that, since Eq.~\eqref{eqn:VUVEqn} is obeyed by the Hadamard tail $V$, we only solve it inside the maximal normal neighbourhood of an arbitrary point $x$. Our $u-v$ plane may be seen as corresponding to the various $\xp$ points given a fixed (arbitrary) point $x$ (the origin $u=v=0$ corresponding to coincidence $x=\xp$).
Therefore, the independent variables should only range over the finite intervals $v\in [0,2\pi)$ and $u\in [0,v]$  (dictated by the range $\gamma\in [0,\pi)$ inside the maximal normal neighbourhood). This is unlike the problem in Refs.~\cite{mark2017recipe,Jonsson:2020npo,PhysRevD.103.124022}, which is for the retarded Green function, which is well-defined for any pair of points anywhere in spacetime, and so with independent variables which are null coordinates in principle each ranging over the whole real line. In order to be able to better map our problem to that in Refs.~\cite{mark2017recipe,Jonsson:2020npo,PhysRevD.103.124022} and the CID problem in Schwarzschild in general,  we shall henceforth consider, without loss of generality, that the PH spacetime points $x$ and $\xp$ have $\theta=\theta'=\pi/2$ and $\varphi'=0$ and, further, that $\gamma\equiv \varphi\in(-\pi,+\pi)$ denotes the azimuthal angle of $x$  (instead of the angular separation, which is in $[0,\pi]$, as until now). The variables $u$ and $v$ continue to be defined as in \eqref{eq:uv} but now with  $\gamma\in (-\pi,+\pi)$ being an azimuthal angle. This means that the region of interest for calculating $V(x,x')$, which is the part of the region of causal separation which lies inside the maximal normal neighbourhood, is bounded by $\eta=2\pi- \gamma$ together with $\eta=\gamma$ if $\gamma\in [0,\pi]$\footnote{\label{ftn:gamma}Here we allow $\gamma$ to take on the value $\pi$ (respectively $-\pi$) so as to refer to the {\it boundary} of the region of interest.}, as explained in Sec.~\ref{sec:PH}, and by $\eta=2\pi+ \gamma$ together with $\eta=-\gamma$ if $\gamma\in [-\pi,0]$\footref{ftn:gamma} after extending the range of $\gamma$ into the negative line. In the CID variables, this means that the region of interest covered by the numerical grid, is given by $u,v\in [0,2\pi)$.

We next describe our scheme and its implementation for obtaining $V$ for various values of $\zeta$. We start by describing a lower order version of the scheme, namely, $\order{h^3}$ of accuracy. We do so in order to more clearly highlight the key distinction in solving the PDE \eqref{eqn:VUVEqn}, containing first-order derivatives, as opposed to the PDE for multipolar  modes of Refs.~\cite{mark2017recipe,Jonsson:2020npo,PhysRevD.103.124022}, not containing any first-order derivatives. We then describe the method at the higher order $\order{h^4}$ of accuracy. In App.~\ref{app:nextOrderScheme} we describe how the scheme can be extended to fifth, and even, sixth order.

\subsubsection{CID scheme setup}

In order to solve the wave Eq.~\eqref{eqn:VUVEqn} with the CID in Eq.~\eqref{eqn:VCIDBC}, we first establish a uniform grid of points on the $u-v$ plane, which we show in Fig.~\ref{fig:CIDGridM2xS2}. The spacing between grid points is $2h$.  The next step is to integrate Eq.~\eqref{eqn:VUVEqn} over an arbitrary $S$-$E$-$N$-$W$ square in the grid (see Fig.~\ref{fig:CIDGridM2xS2}); we seek the value of $V$ at $N$ assuming that its values at $S$, $E$ and $W$ are known. We have

\begin{widetext}
    \begin{align}
        4\int\limits_{SENW} \frac{\partial^2V}{\partial v\partial u} \,\textrm{d}v\,\textrm{d}u+\int\limits_{SENW} Q\frac{\partial V}{\partial v} \,\textrm{d}v\,\textrm{d}u
        +\int\limits_{SENW} S\frac{\partial V}{\partial u} \,\textrm{d}v\,\textrm{d}u+
        \coupl
        \int\limits_{SENW}V \,\textrm{d}v\,\textrm{d}u=0.\label{eqn:squareIntegralV}
    \end{align}
\end{widetext}
Henceforth, a subindex $N$, $E$, $W$, $S$ or $O$ in a quantity means that quantity is to be evaluated at the corresponding point on the grid. The first integral in the left hand side of Eq.~\eqref{eqn:squareIntegralV} can be readily evaluated exactly as
\begin{equation}\label{eqn:firstCIDInt}
    \int\limits_{SENW} \frac{\partial^2V}{\partial v\partial u} \,\textrm{d}v\,\textrm{d}u=V_N-V_E-V_W+V_S,
\end{equation}
where, as mentioned, the subindices $N$, $E$, $W$ and $S$ in $V$ refer to the point on the grid where $V$ is to be evaluated. For the remaining three integrals, we Taylor expand the integrands about the central point $O=(u_O,v_O)$ in the square.
A function $F(v,u)$ which is analytic at the point $O$ admits a Taylor series expansion about $O$:
\begin{widetext}
    \begin{equation}
        F(v,u)=\sum_{\substack{0\leq m,n\leq K\\m+n\leq K}}\frac{1}{m!\, n!}\left(\frac{\partial^{m+n}F}{\partial v^m\partial u^n}\right)_O(v-v_0)^m(u-u_0)^n+\mathcal{O}(h^{K+1}),
        \label{eqn:FGTaylorSeries}
    \end{equation}
\end{widetext}
where  $K$ determines the order in the expansion. We expand in this manner the last three integrands in Eq.~\eqref{eqn:squareIntegralV} (replacing $F$ 
in Eq.~\eqref{eqn:FGTaylorSeries} by the appropriate functions) to the desired order.

\subsubsection{Third order CID scheme}\label{sssec:3OrderScheme}

Using \eqref{eqn:FGTaylorSeries} to order $h^2$, the last three integrals in \eqref{eqn:squareIntegralV} are given by
\begin{align}\label{eqn:squareInts}
    \int\limits_{SENW} Q\frac{\partial V}{\partial v} \,\textrm{d}v\,\textrm{d}u=\,&4h^2Q_O\left(\frac{\partial V}{\partial v}\right)_O+\mathcal{O}(h^4),\\\label{eqn:squareIntsMid}
    \int\limits_{SENW} S\frac{\partial V}{\partial u} \,\textrm{d}v\,\textrm{d}u=\,&4h^2S_O\left(\frac{\partial V}{\partial u}\right)_O+\mathcal{O}(h^4),\\
    \label{eqn:squareIntsEnd}
    \int\limits_{SENW}V \,\textrm{d}v\,\textrm{d}u=\,&4h^2V_O+\mathcal{O}(h^4),
\end{align}
where,  again, the subindex $O$ indicates that the corresponding quantity is evaluated at the point $O$. In this result we only considered the first two leading orders in the Taylor series to obtain the integrals to order $h^3$. However, it can be shown that the contribution to the integrals from the next-to-leading order term in the Taylor series vanishes.

In order to calculate $V$ and its derivatives at the point $O$, we evaluate its Taylor expansion at the points $E,W$ and $S$ to order $h$. This allows us to construct a system of three equations where the unknown variables are $V_O$, $\left(\pd{V}{u}\right)_O$ and $\left(\pd{V}{v}\right)_O$. We find:
\begin{align}\label{eqn:VO2}
    V_O=\,&\frac{V_E+V_W}{2}+\mathcal{O}(h^2),\\
    \left(\frac{\partial V}{\partial u}\right)_O=\,&\frac{V_W-V_S}{2h}+\mathcal{O}(h),\label{eqn:dVduO}\\
    \left(\frac{\partial V}{\partial v}\right)_O=&\,\frac{V_E-V_S}{2h}+\mathcal{O}(h).\label{eqn:dVdvO}
\end{align}

In this way, by using Eqs.~\eqref{eqn:squareInts}--\eqref{eqn:dVdvO} and Eq.~\eqref{eqn:squareIntegralV}, we find that the sought-after value of $V$ at the point $N$ is given by
\begin{widetext}
    \begin{equation}\label{eqn:VNExpression}
        V_N=-V_S-\left(\frac{V_E+V_W-2V_S}{u_O+v_O}-\frac{1}{2}(V_E-V_W)\cot{\frac{v_O-u_O}{2}}\right)h+\left(1
        -\frac{\coupl}{2}
        h^2\right)(V_E+V_W)+\order{h^3},
        \quad \text{for} \quad \gamma_O\neq 0.
    \end{equation}
\end{widetext}
If we compare this expression for $V_N$ with its analogue for the  multipolar modes of the retarded Green function in Schwarzschild in Refs.~\cite{mark2017recipe,Jonsson:2020npo,PhysRevD.103.124022}, we immediately note the additional term linear in $h$ in Eq.~\eqref{eqn:VNExpression}. This term is related with the two {\it first} order derivatives  in Eq.~\eqref{eqn:VUVEqn}, absent in~\cite{mark2017recipe,Jonsson:2020npo,PhysRevD.103.124022}. Furthermore, the order in the right hand side of Eqs.~\eqref{eqn:dVduO}--\eqref{eqn:dVdvO} is connected  with the LTE in Eq.~\eqref{eqn:VNExpression} being {\it third} order (it is easy to check that the coefficient of $h^3$ is not zero), instead of fourth order as in Refs.~\cite{mark2017recipe,Jonsson:2020npo,PhysRevD.103.124022}.

Another key difference between our case and the multipolar-mode analogue appears when $u_O=v_O$ (i.e., along $\gamma=0$), as the term in Eq.~\eqref{eqn:VNExpression} with the $\cot$  function is not well-defined there. Applying the spherical symmetry of $\mt\times\st$, it follows that $V(\eta,\gamma)=V(\eta,-\gamma)$ and so $V_E=V_W$ for any square with a central point such that $u_O=v_O$. Along $\gamma=0$, the  $\cot$ term can thus be calculated by taking the limit $\gamma_O\equiv \frac{v_O-u_O}{2}\to0$ as
\begin{equation}\label{eq:lim VE-VSco}
    \left[(V_E-V_W)\cot\gamma_0\right]_{\gamma_0=0}=\lim_{\gamma_0\to0}\frac{V_E-V_W}{\tan(\gamma_0)}=\order{h}.
\end{equation}
In the last equality we have applied L'H\^{o}pital's rule and expressed $V_E$ and $V_W$  as  small $\gamma$-expansions. With this, it follows that for all squares with $u_O=v_O$, $V_N$ is given by
\begin{align}
    &
    V_N=-V_S-h\left(\frac{V_E-V_S}{v_O}\right)+
    \left(2-\coupl h^2
    \right)V_E+\order{h^3},
\nonumber
    \\&
    \quad \text{for} \quad \gamma_O= 0,
    \label{eqn:VNDiag}
\end{align}
where we have set $V_W=V_E$  due to the spherical symmetry as mentioned above. 

With Eq.~\eqref{eqn:VNDiag} for $V_N$ along $\gamma=0$, we can reformulate the CID problem in the following way. We first split the grid in Fig.~\ref{fig:CIDGridM2xS2} into two triangles, one on each side of the $u=v$ diagonal (i.e., $\gamma=0$). For the choice that we make of applying the scheme to points on the diagonal and {\it bottom}  triangle, $V_N$ in any square with $u_O=v_O$  only depends on $V_E$ and $V_S$ (see Eq.~\eqref{eqn:VNDiag}). This implies that, for the choice of bottom triangle, and after having imposed the symmetry along $\gamma=0$, we only require $V|_{u=0}$ as initial data.  As already pointed out, the values at points in the top triangle can just be obtained using the $V(\eta,\gamma)=V(\eta,-\gamma)$ symmetry. Alternatively, one could choose to apply the scheme to points on the diagonal and the top triangle, in which case we would have the opposite situation: the only required  initial data in that case would be $V|_{v=0}$. Regardless of the choice of triangle, such implementation of the spherical symmetry reduces by (almost) a half the amount of data to calculate.

To summarize, our third order CID scheme consists of  calculating $V_N$ at the  grid points below the $\gamma=0$ line by using Eq.~\eqref{eqn:VNExpression} and at grid points along $\gamma=0$ by using Eq.~\eqref{eqn:VNDiag}. We start by inserting the value of $\nu_0$ into \eqref{eqn:VCIDBC} in order to obtain the CID along $u=0$ (corresponding to our choice of lower triangle). We then evolve the field $V$ for increasing values of $u$ by using   \eqref{eqn:VNExpression} until we reach $\gamma_O=0$;
finally, at $\gamma_O=0$, we switch to using \eqref{eqn:VNDiag}. In the next subsection we develop a scheme of one higher order.

\subsubsection{Fourth order CID scheme}\label{sec:Oh4}

A fourth order CID scheme of course  requires the local errors for the approximations to the integrals in Eq.~\eqref{eqn:squareIntegralV}  to be $\order{h^4}$, as is already the case in Eqs.~\eqref{eqn:squareInts}--\eqref{eqn:squareIntsEnd}. However, expressions \eqref{eqn:dVduO}--\eqref{eqn:dVdvO} for the derivatives needed in \eqref{eqn:squareInts}--\eqref{eqn:squareIntsEnd} lowered the order of the scheme  in the previous subsection to third order. Therefore, in order to obtain a fourth order CID scheme, we  need to include the next higher order in Eqs.~\eqref{eqn:dVduO}--\eqref{eqn:dVdvO}. Including these next higher order terms requires obtaining higher (than in the third order scheme) Taylor coefficients in Eq.~\eqref{eqn:FGTaylorSeries} (with $F=V$).
For instance, we shall see later (see Eqs.~\eqref{eqn:diagDerivIntegrals} and \eqref{eqn:d2Vdgamma2O} below) that obtaining $V_N$ to $\order{h^4}$ along $\gamma=0$ requires the second order derivatives of $V$ at the point $O$. More specifically, for our fourth order CID scheme, we  need to calculate seven more Taylor coefficients in addition to the three given in Eqs.~\eqref{eqn:VO2}--\eqref{eqn:dVdvO}. In order to obtain these coefficients (together with higher order versions of Eqs.~\eqref{eqn:VO2}--\eqref{eqn:dVdvO}) we construct a system of 12 equations as follows. 
We first evaluate Eq.~\eqref{eqn:FGTaylorSeries} (with $F=V$) up to order $h^3$ (inclusive) and its first order derivatives at the points $N$, $E$, $W$ and $S$. Ten of these 12 equations are then used to express $V$ and its first, second and third order derivatives at the point $O$ in terms of $V$ and its first order derivatives at the points $E$, $W$, $N$ and $S$. Out of these ten equations, the following five yield the first and second order derivatives at the point $O$, all of which we shall directly use in our scheme:
\begin{widetext}
    \begin{align}
        \label{eqn:dVduHigherOrder}
        8h\left(\frac{\partial V}{\partial u}\right)_O&=-5V_S-V_E+5V_W+V_N
        -2h\left(\frac{\partial V}{\partial u}
        +\frac{\partial V}{\partial v}\right)_S
        -2h\left(\frac{\partial V}{\partial u}
        -\frac{\partial V}{\partial v}\right)_W+\order{h^4},\\
        \label{eqn:dVdvHigherOrder}
        8h\left(\frac{\partial V}{\partial v}\right)_O&=-5V_S+5V_E+V_W-V_N
        -2h\left(\frac{\partial V}{\partial u}
        +\frac{\partial V}{\partial v}\right)_S
        +2h\left(\frac{\partial V}{\partial u}
        -\frac{\partial V}{\partial v}\right)_E+\order{h^4},\\
        \label{eqn:d2Vdu2O2}
        4h^2\left(\frac{\partial^2V}{\partial u^2}\right)_O&=V_S-V_E-V_W+V_N+2h\left(\frac{\partial V}{\partial u}\right)_E-2h\left(\frac{\partial V}{\partial u}\right)_W+\order{h^4},\\\label{eqn:GdGofEWNSLast}
        4h^2\left(\frac{\partial^2V}{\partial v^2}\right)_O&=V_S-V_E-V_W+V_N-2h\left(\frac{\partial V}{\partial v}\right)_E+2h\left(\frac{\partial V}{\partial v}\right)_W+\order{h^4},\\\label{eqn:d2Vdudv}
        4h^2\left(\frac{\partial^2V}{\partial v\partial u}\right)_O&=V_N+V_S-V_E-V_W+\order{h^4},
    \end{align}
In App.~\ref{app:taylorCoeffs} we give the other  five equations  out of the ten (these other five equations yield  the remaining Taylor coefficients, which we do not directly use  here, as well as a higher order version of Eq.~\eqref{eqn:VO2}; we note that these extra five equations are explicitly needed in the higher order scheme of App.~\ref{app:nextOrderScheme}).
The remaining two equations out of the total of twelve are used to calculate the first order derivatives of $V$ at the point $N$:
    \begin{align}
        \label{eq:dV/du,v h3}
        \left(\pd{V}{u}\right)_N=\,&\frac{V_S-V_E-V_W+V_N}{h}-\left(\pd{V}{u}\right)_E+\left(\pd{V}{u}\right)_W+\left(\pd{V}{u}\right)_S+\order{h^3},\\\notag
        \left(\pd{V}{v}\right)_N=\,&\frac{V_S-V_E-V_W+V_N}{h}+\left(\pd{V}{v}\right)_E-\left(\pd{V}{v}\right)_W+\left(\pd{V}{v}\right)_S+\order{h^3}.
    \end{align}
\end{widetext}
The values of these two derivatives at the point $N$ together with the value of $V_N$ will be required at the next step while we move rightwards on the $u-v$ plane, since the current point $N$ will then become the point $W$ at the next step. In principle, we now have all the necessary equations to construct a fourth order CID scheme for $V_N$. However, we note the following. Eq.~\eqref{eqn:VO2} for $V_O$ is to order $h^2$ while Eqs.~\eqref{eqn:dVduHigherOrder}--\eqref{eqn:dVdvHigherOrder} for the first order derivatives at $O$ are to order $h^3$. 
We have found that our  numerical results  are more accurate if, instead of using Eqs.~\eqref{eqn:dVduHigherOrder}--\eqref{eqn:dVdvHigherOrder}, we  use the following expressions\footnote{For obtaining Eqs.~\eqref{eqn:dVduO2}--\eqref{eqn:dVdvO2} we constructed a system of 4 equations --other than the system of 12 equations mentioned above-- by using Eq.~\eqref{eqn:FGTaylorSeries} evaluated at the $E$, $W$, $N$ and $S$ points. We then solved this system for $V_O$, $\evalO{\pd{^2V}{u\partial v}}$, $\evalO{\pd{V}{u}}$, and $\evalO{\pd{V}{v}}$, and Eqs.~\eqref{eqn:dVduO2}--\eqref{eqn:dVdvO2} are the expressions for the  two latter quantities.} for these derivatives to order $h^2$:
\begin{align}\label{eqn:dVduO2}
    4h\evalO{\pd{V}{u}}=\,&V_W-V_E-V_S+V_N+\order{h^3},\\
    \label{eqn:dVdvO2}
    4h\evalO{\pd{V}{v}}=\,&V_E-V_W-V_S+V_N+\order{h^3}.
\end{align}
We now have a system with all the necessary equations
for squares with $u_O\neq v_O$ (i.e., $\gamma_0\neq 0$; we treat the case $\gamma_O=0$ separately below). We simply  put Eqs.~\eqref{eqn:VO2}, \eqref{eqn:dVduO2}--\eqref{eqn:dVdvO2} and \eqref{eqn:squareInts}--\eqref{eqn:squareIntsEnd} back into Eq.~\eqref{eqn:squareIntegralV} and isolate for $V_N$. This yields
    \begin{widetext}
    \begin{align}\notag
        &V_N=V_E+V_W-V_S-\left(\frac{V_E+V_W-2V_S}{u_O+v_O}-\frac{1}{2}(V_E-V_W)\cot\frac{v_O-u_O}{2}\right)h\\\label{eq:CN 4thO}
        &+\left(\frac{(2-(u_O+v_O)^2\zeta)(V_E+V_W)-4V_S}{2(u_O+v_O)^2}-\frac{V_E-V_W}{2(u_O+v_O)}\cot\frac{v_O-u_O}{2}\right)\cdot\left(h^2-\frac{h^3}{u_O+v_O}\right)+\order{h^4}, \quad \gamma_O\neq 0.
    \end{align}
\end{widetext}

Similarly to the third order scheme in the previous subsection, there is a delicate cancellation in Eq.~\eqref{eq:CN 4thO} at $u_O=v_O$ (i.e., at $\gamma_O=0$), as can be seen by the terms with the $\cot$ function.
Unfortunately, the expression  in Eq.~\eqref{eq:lim VE-VSco} for the limit to the line of symmetry $\gamma=0$ is no longer good enough since its $\order{h}$ term cannot be ignored in a fourth order scheme. 
In order to resolve this, we rewrite the sum of the second and third integrals in Eq.~\eqref{eqn:squareIntegralV} as
\begin{widetext}
    \begin{align}
        \int\limits_{SENW}\left(Q\frac{\partial V}{\partial v} +S\frac{\partial V}{\partial u}\right)\textrm{d}u\,\textrm{d}v&=
        \frac{8h^2}{u_O+v_O}\left(\frac{\partial V}{\partial v}+\frac{\partial V}{\partial u}\right)_O-4h^2\cot{\frac{v_O-u_O}{2}}\left(\frac{\partial V}{\partial v}-\frac{\partial V}{\partial u}\right)_O+\order{h^4}\nonumber\\
        &=\frac{8h^2}{u_O+v_O}\left(\frac{\partial V}{\partial v}+\frac{\partial V}{\partial u}\right)_O-4h^2\cot{\gamma_O}\left(\pd{V}{\gamma}\right)_O+\order{h^4}.\label{eq:2nd+3rd as gamma}
    \end{align}
\end{widetext}
We have written the second term on the right hand side of \eqref{eq:2nd+3rd as gamma} in terms of $\gamma$ instead of $u$ and $v$ so that we can more easily take the desired limit $\gamma_O\to 0$. Taking the limit $\gamma_O\to0$  in the first term on the right hand side of \eqref{eq:2nd+3rd as gamma}  is equivalent to setting $u_O=v_O$ and $\evalO{\pd{V}{u}}=\evalO{\pd{V}{v}}$, as a consequence  of the symmetries of $\mt\times\st$. For the limit $\gamma_O\to0$ in the second term, we apply again  L'H\^{o}pital's rule\footnote{The $V(\eta,\gamma)=V(\eta,-\gamma)$ symmetry implies $\left(\pd{V}{\gamma}\right)_{\gamma=0}=0$, which prompts us to use L'H\^{o}pital's rule.}. This yields,
\begin{align}
\notag
    &\lim_{\gamma_O\to0}\,\,\int\limits_{SENW}\left(Q\frac{\partial V}{\partial v} +S\frac{\partial V}{\partial u} \right)\textrm{d}v\,\textrm{d}u=\\\label{eqn:diagDerivIntegrals}
    &\,\frac{8h^2}{v_O}\left(\frac{\partial V}{\partial v}\right)_O-4h^2\left(\pd{^2V}{\gamma^2}\right)_O+\order{h^4},
\ \text{for}\ \gamma_O=0.
\end{align}
When using $u$ and $v$ (instead of $\eta$ and $\gamma$) as independent variables, the  second order derivative with respect to $\gamma$ appearing above can be written as
\begin{equation}\label{eqn:d2Vdgamma2O}
    \left(\pd{^2V}{\gamma^2}\right)_O=\left(\pd{^2V}{u^2}\right)_O-2\left(\pd{^2V}{u\partial v}\right)_O+\left(\pd{^2V}{v^2}\right)_O.
\end{equation}

Although in principle these second order derivatives are only required along $\gamma=0$, in order to know their values there, we first need to know their values away from $\gamma=0$, so that we can then propagate them to  $\gamma=0$.

We now have a system with all the necessary equations for the $u_O=v_O$ (i.e., $\gamma_O=0$) case and proceed to solve it and evaluate  the required integrals in Eq.~\eqref{eqn:squareIntegralV}. The first and fourth integrals in Eq.~\eqref{eqn:squareIntegralV} are the same as in the third order scheme, namely, Eqs.~\eqref{eqn:firstCIDInt} and \eqref{eqn:squareIntsEnd}, respectively. In their turn, the second and third integrals  in Eq.~\eqref{eqn:squareIntegralV} are now calculated using Eq.~\eqref{eqn:diagDerivIntegrals}--\eqref{eqn:d2Vdgamma2O} together with the Taylor coefficients calculated via Eqs.~\eqref{eqn:d2Vdu2O2}--\eqref{eqn:d2Vdudv} and \eqref{eqn:dVdvO2}. Isolating for $V_N$ in the resulting Eq.~\eqref{eqn:squareIntegralV} then  yields

\begin{widetext}
    \begin{align}\notag
        V_N=&\,2V_E-V_S+\left[\left(\pd{V}{v}-\pd{V}{u}\right)_E+\frac{V_S-V_E}{v_O}\right]h\\\label{eq:CN 4thO,gamma=0}
        &\,+\frac{1}{2v_O}\left[\frac{V_S-(1-2\zeta{v_O}^2)V_E}{v_O}+\left(\pd{V}{v}-\pd{V}{u}\right)_E\right]\cdot\left(h^2+\frac{h^3}{2v_O}\right)+\order{h^4},\quad \gamma_O=0.
    \end{align}
\end{widetext}
In order to write this equation in terms of data  on the $E$ and $S$ points only, we have used $V_W=V_E$ in addition to $\left(\pd{V}{u}\right)_W=\left(\pd{V}{v}\right)_E$ and $\left(\pd{V}{v}\right)_W=\left(\pd{V}{u}\right)_E$, all of which are valid for $\gamma_O=0$ as follows from the $V(\eta,\gamma)=V(\eta,-\gamma)$ symmetry. In this way, we are still able to evolve (almost) half the grid points similarly to what we did in the third order CID scheme.

Clearly, our fourth order CID scheme requires calculating data on the light cone other than the data for $\hat{V}$. Specifically, Eqs.~\eqref{eqn:dVduHigherOrder}--\eqref{eq:dV/du,v h3} require the first order derivatives of $V$ on the lightcone. This calculation is easily achieved by differentiating Eq.~\eqref{eqn:VsigmaExpansion} once with respect to $u$ and once with respect to $v$  and evaluating the derivatives at $\sigma=-\frac{1}{2}uv=0$. We thus readily obtain these derivatives in terms of the Hadamard coefficients $\nu_0$ and $\nu_1$:
\begin{align}\label{eqn:addVCIDBC}
    \left.\frac{\partial V}{\partial u}\right|_{u=0}=\,&-\frac{1}{2}{\nu_0}'\left(\frac{v}{2}\right)-\frac{v}{2}\nu_1\left(\frac{v}{2}\right),\\
    \left.\frac{\partial V}{\partial u}\right|_{v=0}=\,&-\frac{1}{2}{\nu_0}'\left(-\frac{u}{2}\right),\\
    \left.\frac{\partial V}{\partial v}\right|_{u=0}=\,&\frac{1}{2}{\nu_0}'\left(\frac{v}{2}\right),\\\label{eqn:addVCIDBCEnd}
    \left.\frac{\partial V}{\partial v}\right|_{v=0}=\,&\frac{1}{2}{\nu_0}'\left(-\frac{u}{2}\right)-\frac{u}{2}\nu_1\left(-\frac{u}{2}\right),
\end{align}
where the primes indicate differentiation with respect to $\gamma$. We note that, as it should be in a characteristic initial value problem like this one, this data for the derivatives on the light cone is obtainable from the value of $V$ on the light cone (i.e., $\hat{V}$). Indeed, in the case here of PH, it is $\nu_0=\hat{V}$ and  $\nu_1$ can be uniquely obtained from $\nu_0$ together with the condition of regularity at coincidence (see Eq.~(24) in \cite{casals-nolan-global-hadamard}; in fact, a Hadamard coefficient $\nu_n$, for any $n>0$, can be similarly obtained from the previous coefficient $\nu_{n-1}$ in a generic spacetime -- see~\cite{DeWitt:1960,Decanini:Folacci:2005a,Ottewill:2009uj}).

The fourth order CID scheme is then established as follows: We construct the usual uniform  grid (see Fig.~\ref{fig:CIDGridM2xS2}) and place the initial data ($V$ and its first order derivatives) along $u=0$ if we want to evolve points on and below the $\gamma=0$ line\footnote{For the opposite choice of evolving points on and {\it above} the $\gamma=0$ line, the initial data is instead  placed along $v=0$.}. Then, we use Eq.~\eqref{eq:CN 4thO} to calculate $V_N$ for squares not centered along $\gamma=0$. For squares centered along $\gamma=0$, $V_N$ is instead given by Eq.~\eqref{eq:CN 4thO,gamma=0}. Finally, we use Eq.~\eqref{eq:dV/du,v h3} in order to calculate the first order derivatives (both for $\gamma=0$ and $\gamma \neq 0$). We note that the issue with $\gamma=0$  does not appear in \cite{Lousto:1997wf,mark2017recipe,Jonsson:2020npo,PhysRevD.103.124022} since in those cases the PDE does not have a singular point like the $\gamma=0$ point here for the {\it full} wave equation in $\mt\times\st$.

After having provided complete prescriptions for the CID scheme to two different orders (namely, third and fourth orders), we implemented these using the computer algebra software {\it Mathematica}. In the following subsection we show our results  for $V$, including comparisons  against previous results obtained using different approaches to calculate $V$ (see \cite{Casals:2012px}) as well as the presentation of  new results for $V$.

\subsection{Results for $V$}\label{sec:results V}

\begin{figure}
    \centering
    \includegraphics[scale=0.65]{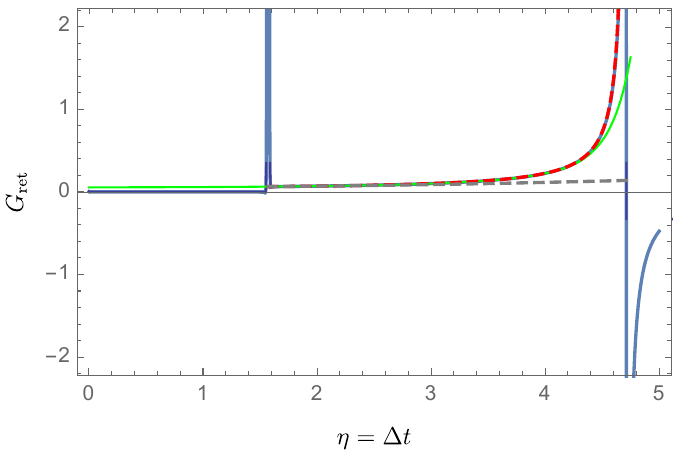}\\~\\
    \includegraphics[width=0.45\textwidth]{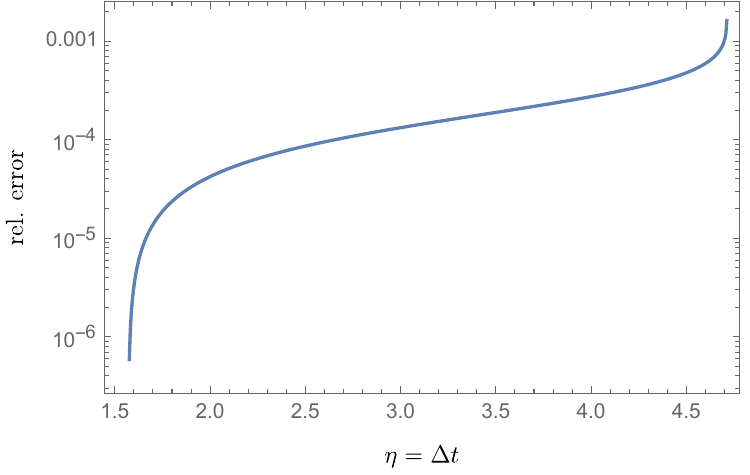}
    
    \caption{Quantities  as functions of $\eta=\Delta t$ in $\mt\times\st$ with $\coupl=1/4$, $\Delta y=y-y'=0$ and $\gamma=\pi/2$. Top plot: Retarded Green function as an $\ell$-mode sum (blue) and $V$ using the fourth order CID scheme (dashed red), a small coordinate-separation expansion (green) and approximated by $\nu_0+\nu_1\sigma$ (dashed gray). Bottom plot: Relative error between the third order and fourth order schemes for calculating $V$ with $h=0.00261799$.}
    \label{fig:GretComparison}
\end{figure}

In this section we show our results for the Hadamard biscalar $V(x,x')$. In the top plot of Fig.~\ref{fig:GretComparison} we consider the case of $x$ and $x'$ on static paths with $y=y'$ and $\gamma=\pi/2$ for  $\coupl=1/4$. In it, we compare the following: $V$ obtained using the fourth order CID scheme of Sec.~\ref{sec:Oh4} with the choice of $h=0.00261799$ (in dashed red); the retarded Green function $G_{\textrm{ret}}$ calculated with the multipolar $\ell$-mode sum (capped at the finite value $\ell=800$) expression given in Eq.~(134) of Ref.~\cite{Casals:2012px}\footnote{We note that in the last expression in Eq.~(134) of Ref.~\cite{Casals:2012px} there is a  missing factor $\theta(-\sigma_{\mathbb{M}_2})$.} (in blue); the crude approximation $\nu_0+\nu_1\sigma$ to $V$ (in dashed gray) from Eq.~\eqref{eqn:VsigmaExpansion}; $V$ calculated using a small coordinate distance expansion (see Ref.~\cite{CDOWb}) (in green).  The first divergence of $G_{\textrm{ret}}$ in the top plot of Fig.~\ref{fig:GretComparison}    is at $\eta=\pi/2$, corresponding to the direct null geodesic divergence $\delta(\sigma)$ as per the Hadamard form Eq.~\eqref{grhad}. This divergence signals the start of causal separation. As explained in Sec.~\ref{sec:PH}, $V$  and $G_{\textrm{ret}}$ should agree in the region between this divergence and the next divergence at $\eta=3\pi/2$, which signals the end of the maximal normal neighbourhood and corresponds to a null geodesic having crossed a caustic at $\gamma=\pi$. Thus, this latter divergence should be of type $\text{PV}\left(1/\sigma\right)$, in agreement with the plot.

The top plot of Fig.~\ref{fig:GretComparison}  also shows that the CID scheme has good agreement with the $\ell$-mode-sum calculated $G_\textrm{ret}$. Indeed, in the bottom plot of Fig.~\ref{fig:GretComparison} we show that the relative error between the two CID schemes, with LTEs of orders $\order{h^3}$ and $\order{h^4}$, with the same stepsize $2h=2\times0.00261799$, is at least of order  $10^{-4}$. Let us check this value for consistency. Let $e_3$ and $e_4$ denote the {\it global} truncation errors (GTEs) for the schemes with {\it local} truncation errors $\mathcal{O}(h^3)$ and $\mathcal{O}(h^4)$, respectively. For the $n$-th  evolved point in the grid, these errors are given by $e_3=\order{n(2h)^3}$ and $e_4=\order{n(2h)^4}$. For the computation of $V$ at a fixed point close to the end of the normal neighbourhood, about $n=\order{(2h)^{-2}}$ evaluations\footnote{More accurately, it should be  $n=\order{(2h)^{-2}/2}$, where the extra factor $1/2$ arises due to the fact that the grid, which in principle would be a square as per Fig.~\ref{fig:CIDGridM2xS2}, really becomes a triangle because of application of the symmetries mentioned around Eq.~\eqref{eqn:VNDiag}. But $n=\order{(2h)^{-2}}$ is fine as an order of magnitude.}  are carried out, yielding GTEs $e_3=\order{2h}$ and $e_4=\order{(2h)^{2}}$. Thus, for a point close to the end of the normal neighbourhood  in the case of Fig.~\ref{fig:GretComparison}, we have $e_3=\order{10^{-3}}$ and $e_4=\order{10^{-5}}$, which can be taken as relative errors since $V=\order{1}$ close the end of the normal neighbourhood. The GTE $e_3=\order{10^{-3}}$ is consistent with the relative error between the two schemes shown in the bottom plot of Fig.~\ref{fig:GretComparison}. On its own, this relative error does not tell us much about the improvement that we achieve by applying a higher order scheme. Instead, we look at the GTE. Given three approximations to $V$ calculated with the same  CID scheme but with three different stepsizes ($2h$, $4h$ and $8h$ in our case, with corresponding solutions denoted by $V_{(2h)}$, $V_{(4h)}$ and $V_{(8h)}$), one can verify that (see, e.g., \cite{McDonoughNumericalAnalysis})
\begin{equation}\label{eqn:GTEEqn}
    \frac{V_{(2h)}-V_{(4h)}}{V_{(4h)}-V_{(8h)}}=\frac{(2h)^k-(4h)^k}{(4h)^k-(8h)^k}+\order{h}=\frac{1}{2^k}+\order{h},
\end{equation}
where $k$ denotes the  order of the  GTE  of the scheme. As justified above, we  expect $k=1$  for a third order CID scheme and $k=2$ for a fourth order one.

\begin{figure}
    \centering
    \includegraphics[width=0.45\textwidth]{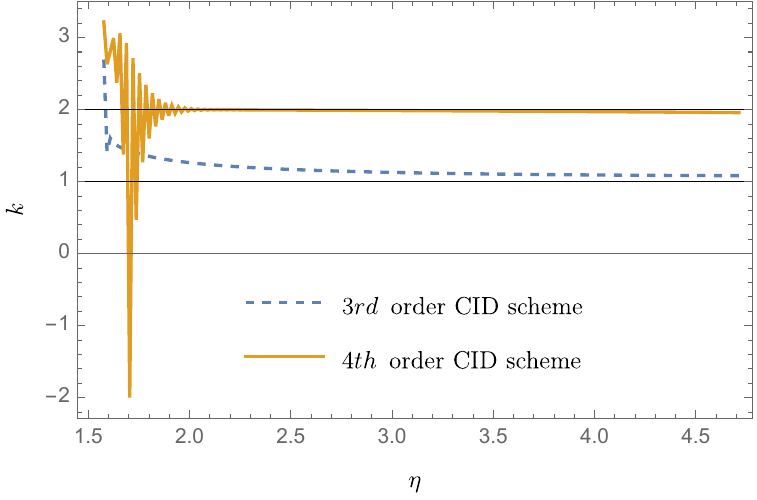}\\~\\
    \includegraphics[width=0.45\textwidth]{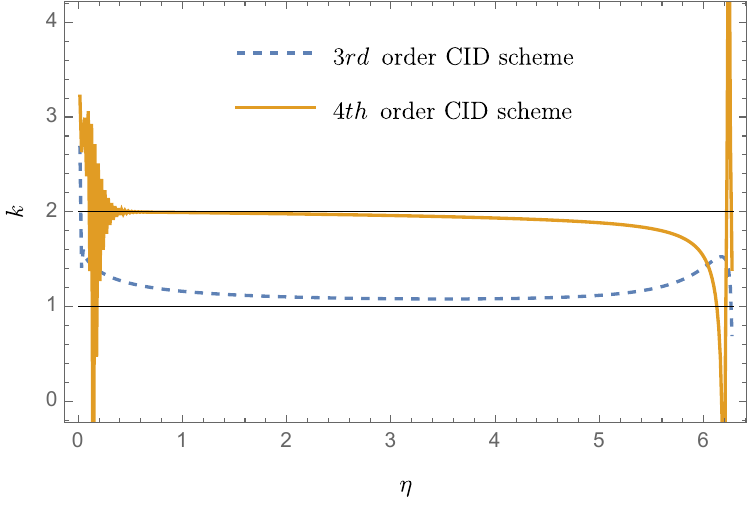}
    \caption{GTE order $k$ for the third and fourth order CID schemes for $\gamma=1.56$ (top) and $\gamma=0$ (bottom). The expected values for the order $k$ are close to the theoretical values (namely, $k=1$ for the third order scheme and $k=2$ for the fourth order one, drawn as horizontal black continuous lines).}
    \label{fig:GTEPlots}
\end{figure}

In Fig.~\ref{fig:GTEPlots} we plot $k$ in Eq.~\eqref{eqn:GTEEqn} as a function of $\eta$ for two different fixed values of the angle:  $\gamma=1.56$ in the top plot and $\gamma=0$ in the bottom one. These two plots show the consistency between the numerically-obtained values of $k$ and the theoretically-expected values of $k$ (namely, $k=1$ and $k=2$ for the third and fourth order schemes, respectively). There is some noise near $\eta=\gamma$ in both plots, which are nevertheless oscillations about the expected value of $k$, which is merely meant to be approximated by \eqref{eqn:GTEEqn}; there is also some oscillations at the end of the lower plot (i.e., approaching the end of the maximal normal neighborhood), which is unsurprising since that plot is for $\gamma=0$ and its end is at $\eta= 2\pi-\gamma=2\pi$, so a caustic point (caustics are at $\eta=\gamma=m\, \pi$, for $m\in\mathbb{Z}$, and it is expected that there is  an enhancement in the singularity of the retarded Green function at caustics, as that is the case in Schwarzschild~\cite{casals2016global}).

While the value of $V$ in Fig.~\ref{fig:GretComparison} had already been obtained before (namely, in~\cite{Casals:2012px}), in the next plots we present new values of $V$ in PH.

\begin{figure}
    \centering
    \includegraphics[scale=0.5]{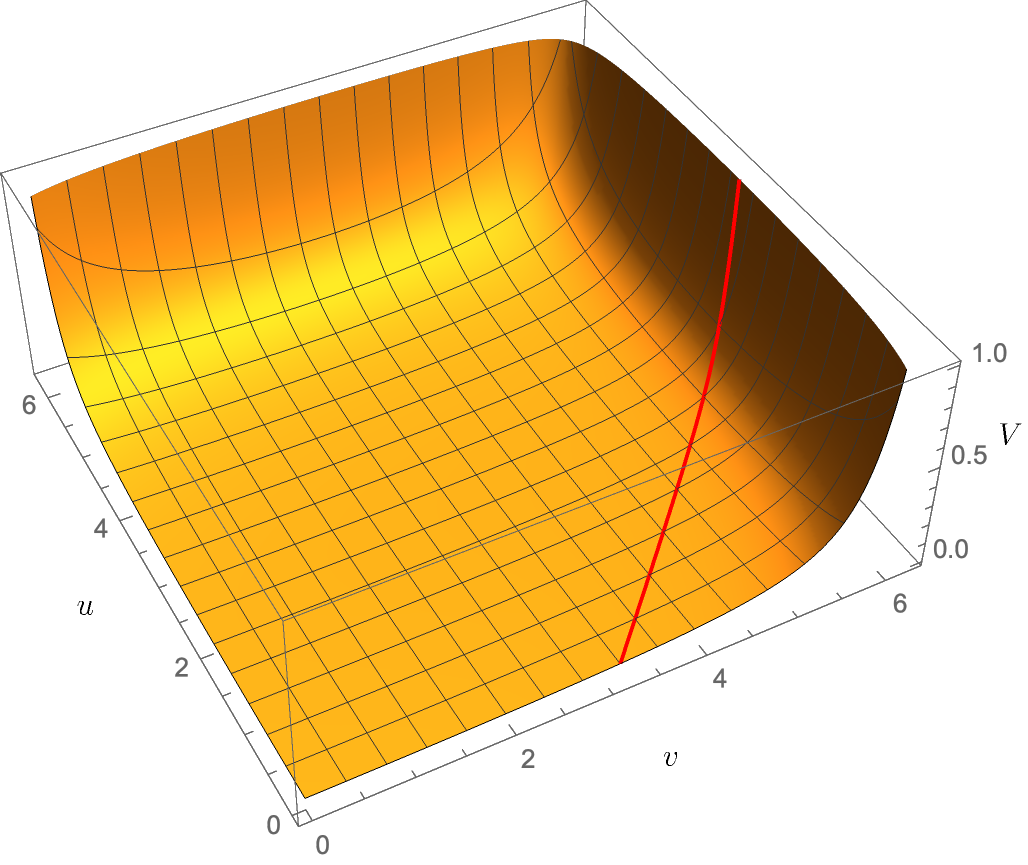}
    \caption{Plot of $V$ for
    all pairs of points in normal neighbourhoods with
    $\coupl=1/4$. The $u=2\pi$ and $v=2\pi$ lines correspond to the end of the normal neighbourhood where the (leading) singularity of $G_\textrm{ret}$, and so of $V$, is of type $\text{PV}\left(1/\sigma\right)$ (when away from caustics). The red line is along the static worldline considered in Fig.~\ref{fig:GretComparison}.}
    \label{fig:V3D}
\end{figure}

In Fig.~\ref{fig:V3D} we show the  plot of $V$ (obtained with the fourth order scheme) for $\zeta=1/4$ for {\it any} pair of spacetime points (as long as they lie in normal neighbourhoods, so that $V$ is defined). The red line corresponds to the static worldlines  of Fig.~\ref{fig:GretComparison} (namely, $y=y'$ and $\gamma=\pi/2$). Evolving  CID  has allowed us to calculate $V$  {\it everywhere} where it is defined.

\begin{figure}
    \vspace{1cm}
    \centering
    \includegraphics[width=0.45\textwidth]{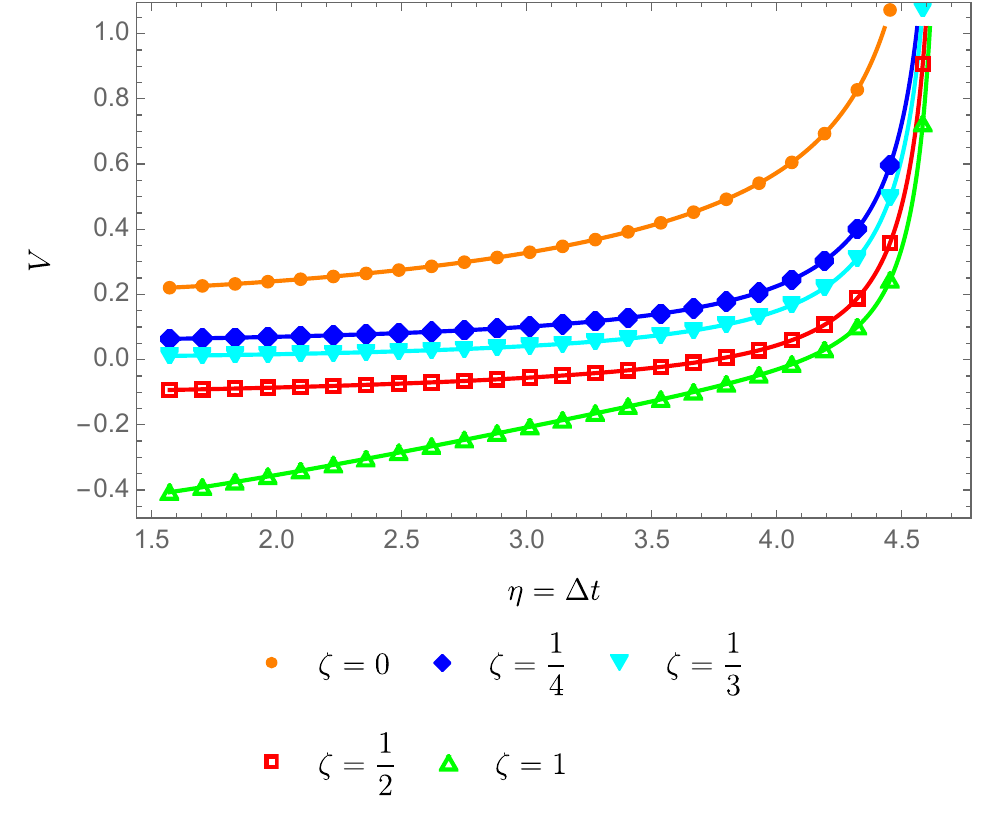}\\~\\
    \includegraphics[width=0.45\textwidth]{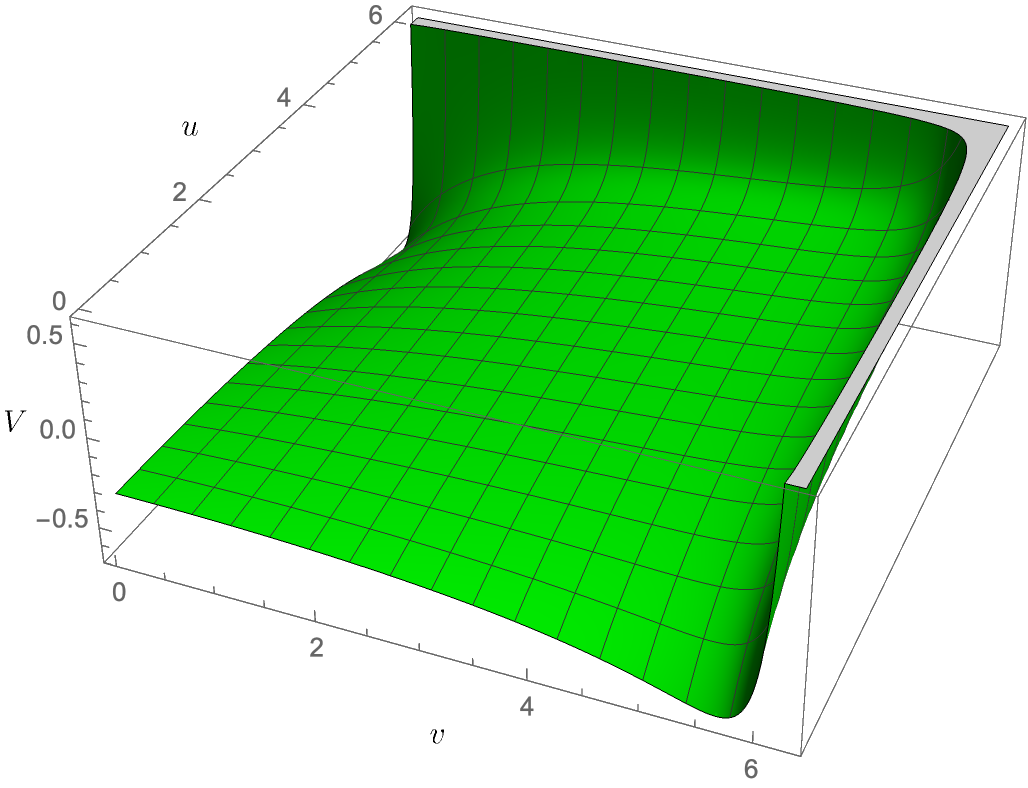}
    \caption{Top: Plot of $V$ as a function of $\eta$ for different values of $\zeta$ and for the same scenario (and stepsize) as in Fig.~\ref{fig:GretComparison}. Bottom: 3D plot of $V$ as a function of $u$ and $v$ for $\zeta=1$.}
    \label{fig:VXi0468}
\end{figure}

We also calculated $V$ with the third order scheme for various   values of $\zeta\neq 1/4$. In the top plot of Fig.~\ref{fig:VXi0468} we show these results for the same worldline as in Fig.~\ref{fig:GretComparison}. For this particular worldline (which has $\gamma=\pi/2$), the magnitude of $V$ decreases as $\zeta$ increases. In the bottom plot of Fig.~\ref{fig:VXi0468} we again plot $V$ for all possible pairs of spacetime points, but now  for $\zeta=1$. We can see in it that there is a more marked change in the form, with respect to Fig.~\ref{fig:V3D} for $\zeta=1/4$,  near the caustic $\gamma=\pm\pi$. We note that we also calculated $V$ for all pairs of points for the other values of $\zeta$ that we used in the top plot of Fig.~\ref{fig:VXi0468} but we do not display the full results since their behaviour was not so different from that in Fig.~\ref{fig:V3D} for $\zeta=1/4$. 

\section{Discussion}\label{sec:Discussion}

In this work we have presented and implemented a new method for calculating the Hadamard tail biscalar $V(x,x')$ for wave propagation on curved background spacetimes where null geodesics cross. This method consists of integrating the homogeneous wave equation using Characteristic Initial Data on the light cone. We have provided a proof-of-concept for this method  by applying it to PH, $\mt\times\st$. Furthermore, we have calculated $V$ for new cases: at {\it all} spacetime points where it is defined, for various values of $\coupl\equiv m^2+2\xi$. The calculation of $V$ (and of the retarded Green function) at all pairs of points is useful, in particular, for a potential application to the self-consistent orbital evolution of a particle via the self-force.

The calculation in $\mt\times\st$ is technically easier than in black hole spacetimes: First, the Characteristic Initial Data $\left.V(x,x')\right|_{\sigma=0}$  is known analytically and, second, the wave equation is reduced to a {\it two}-dimensional PDE. As for the first point, we note that $V(x,x')$ was numerically calculated  along null geodesics in Schwarzschild in~\cite{Ottewill:2009uj} by solving transport equations, and thus its Characteristic Initial Data in Schwarzschild is readily available, while~\cite{Ottewill:2009uj} also provides a prescription for its calculation in Kerr. 
As for the second point, the wave equation in Schwarzschild would acquire an extra dimension, thus becoming a {\it three}-dimensional PDE, for which there exist numerical techniques. Furthermore, the three-dimensional PDE in Schwarzschild contains first-order derivatives (at least through the angular part),  which the scheme presented here in the context of PH has provided a way of dealing with. In Kerr, the PDE would become {\it four}-dimensional, entailing a greater numerical challenge. We intend to undertake the calculation of $V(x,x')$ in these black hole spacetimes in the future.


\section{{Acknowledgments.}}
We are grateful to Adrian Ottewill and Barry Wardell for useful discussions. 
We also thank the anonymous referee for helping us improve the paper.
 M.C.\ acknowledges partial financial support by CNPq (Brazil), process number 314824/2020-0. D. Q. A. acknowledges support from FAPERJ (process number 200.804/2019) and CNPq (process number 140951/2017-2).


\appendix

\section{Remaining equations for Taylor coefficients in the 4th order scheme}\label{app:taylorCoeffs}

As pointed out in Sec.~\ref{sec:Oh4}, in order to obtain all the twelve equations needed for the fourth order scheme, it is necessary to calculate five Taylor coefficients in Eq.~\eqref{eqn:FGTaylorSeries} additional to those given in Eqs.~\eqref{eqn:dVduHigherOrder}--\eqref{eq:dV/du,v h3}. The expressions for the five remaining Taylor coefficients are: 
\begin{widetext}
    \begin{align}\label{eqn:VOhigherOrder}
        4V_O&=2V_E+2V_W+h\left(\frac{\partial V}{\partial u}-\frac{\partial V}{\partial v}\right)_E-h\left(\frac{\partial V}{\partial u}-\frac{\partial V}{\partial v}\right)_W+\order{h^4},\\
        \frac{2}{3}h^3\left(\frac{\partial^3V}{\partial v^3}\right)_O&=V_S-V_E+h\left(\frac{\partial V}{\partial v}\right)_S+h\left(\frac{\partial V}{\partial v}\right)_E+\order{h^4},\\ \frac{2}{3}h^3\left(\frac{\partial^3V}{\partial u^3}\right)_O&=V_S-V_W+h\left(\frac{\partial V}{\partial u}\right)_S+h\left(\frac{\partial V}{\partial u}\right)_W+\order{h^4},\\
        4h^3\left(\frac{\partial^3V}{\partial v^2\partial u}\right)_O&=V_N+V_S-V_E-V_W+2h\left(\frac{\partial V}{\partial v}\right)_S-2h\left(\frac{\partial V}{\partial v}\right)_W+\order{h^4},\\\label{eqn:GdGofEWNSNLast}
        4h^3\left(\frac{\partial^3V}{\partial v\partial u^2}\right)_O&=V_N+V_S-V_E-V_W+2h\left(\frac{\partial V}{\partial u}\right)_S-2h\left(\frac{\partial V}{\partial u}\right)_E+\order{h^4}.
    \end{align}
\end{widetext}
The expressions above, apart from having been derived as part of  the set of equations in the fourth order scheme, will be directly used in the higher order schemes described in the next appendix.

\section{Going beyond a fourth order CID scheme}\label{app:nextOrderScheme}\label{sec:Higher order}

In this appendix we lay out the ground work for deriving a fifth --and potentially, sixth-- order scheme, by following a similar prescription to that described in Secs.~\ref{sssec:3OrderScheme} and \ref{sec:Oh4}. For a fifth or sixth order CID scheme, it is necessary to include the next nontrivial order  in Eqs.~\eqref{eqn:squareInts}--\eqref{eqn:squareIntsEnd}. The corresponding equations are given by:
\begin{widetext}
    \begin{align}\notag
        \int\limits_{SENW} Q\frac{\partial V}{\partial v} \,\textrm{d}v\,\textrm{d}u=\,&4h^2Q_O\left(\frac{\partial V}{\partial v}\right)_O+\frac{2}{3}\left[2\evalO{\pd{Q}{u}\pd{^2V}{u\partial v}}+\evalO{\pd{V}{v}}\evalO{\pd{^2Q}{u^2}+\pd{^2Q}{v^2}}\right.+\\\label{eqn:higherOrderIntegral}
        &\left.2\evalO{\pd{Q}{v}\pd{^2Q}{v^2}}+Q_O\evalO{\pd{^3V}{u^2\partial v}+\pd{^3V}{v^3}}\right]h^4+\mathcal{O}(h^6),\\\notag
        \int\limits_{SENW} S\frac{\partial V}{\partial u} \,\textrm{d}v\,\textrm{d}u=\,&4h^2S_O\left(\frac{\partial V}{\partial u}\right)_O+\frac{2}{3}\left[2\evalO{\pd{S}{v}\pd{^2V}{u\partial v}}+\evalO{\pd{V}{u}}\evalO{\pd{^2S}{u^2}+\pd{^2S}{v^2}}\right.+\\
        &\left.2\evalO{\pd{S}{u}\pd{^2V}{u^2}}+S_O\evalO{\pd{^3V}{u^3}+\pd{^3V}{u\partial v^2}}\right]h^4+\mathcal{O}(h^6),\\\label{eqn:higherOrderIntegralEnd}
        \int\limits_{SENW}V \,\textrm{d}v\,\textrm{d}u=\,&4h^2V_O+\frac{2}{3}\left(\pd{^2V}{u^2}+\pd{^2V}{v^2}\right)_O h^4+\mathcal{O}(h^6),
    \end{align}
\end{widetext}
where, as usual, the subscript $O$ on  a quantity in brackets indicates that it is evaluated at the point $O$.

When replacing the expressions in Eq.~\eqref{eqn:firstCIDInt} and Eqs.~\eqref{eqn:higherOrderIntegral}--\eqref{eqn:higherOrderIntegralEnd} back into Eq.~\eqref{eqn:squareIntegralV} and isolate for $V_N$, we find
\begin{widetext}
    \begin{align}
    \notag
        V_N=\,&V_E+V_W-V_S-\left[Q_O\evalO{\pd{V}{v}}+S_O\evalO{\pd{V}{u}}+
        \coupl\,
        V_O\right]h^2-\\\notag
        &\frac{1}{6}\left[
        \coupl\,
        \evalO{\pd{{}^2V}{u^2}+\pd{{}^2V}{v^2}}+2\evalO{\pd{Q}{v}}\evalO{\pd{^2V}{v^2}}+2\evalO{\pd{S}{u}}\evalO{\pd{^2V}{u^2}}\right.+\\\notag
        &\evalO{\pd{V}{v}}\evalO{\pd{^2Q}{u^2}+\pd{^2Q}{v^2}}+2\evalO{\pd{^2V}{u\partial v}}\evalO{\pd{Q}{u}+\pd{S}{v}}+\evalO{\pd{V}{u}}\evalO{\pd{^2S}{u^2}+\pd{^2S}{v^2}}+\\\label{eq:Vn highO}
        &\left.S_O\evalO{\pd{{}^3V}{u\partial v^2}+\pd{^3V}{u^3}}+Q_O\evalO{\pd{{}^3V}{u^2\partial v}+\pd{^3V}{v^3}}\right]h^4+\order{h^6}.
    \end{align}
\end{widetext}
The next step is to replace into Eq.~\eqref{eq:Vn highO} the expressions for the Taylor coefficients given in  Eqs.~\eqref{eqn:dVduHigherOrder}--\eqref{eqn:d2Vdudv} and Eqs.~\eqref{eqn:VOhigherOrder}--\eqref{eqn:GdGofEWNSNLast} and then isolate for $V_N$. We do not display the resulting expression for $V_N$ since it is very long and trivial to obtain. We note that, despite Eq.~\eqref{eq:Vn highO} being $\order{h^6}$, the expressions for the Taylor coefficients given in  Eqs.~\eqref{eqn:dVduHigherOrder}--\eqref{eqn:d2Vdudv} and Eqs.~\eqref{eqn:VOhigherOrder}--\eqref{eqn:GdGofEWNSNLast} reduce the order of the resulting  expression of $V_N$ by one (i.e., to $\order{h^5}$). Thus, we end up with a fifth order CID scheme. One could of course use Eq.~\eqref{eq:Vn highO} for a sixth order scheme merely by obtaining expansions for the Taylor coefficients appearing in it to one order higher than in Eqs.~\eqref{eqn:dVduHigherOrder}--\eqref{eqn:d2Vdudv} and Eqs.~\eqref{eqn:VOhigherOrder}--\eqref{eqn:GdGofEWNSNLast}.

Finally, and similarly to the third and fourth order CID schemes in  Secs.~\ref{sssec:3OrderScheme} and \ref{sec:Oh4}, the $u_O=v_O$ (i.e., $\gamma_O=0$) case should be handled separately. The coefficient of $h^4$ in Eq.~\eqref{eq:Vn highO} involves derivatives of the functions $Q$ and $S$ evaluated at the point $O$. As a consequence, terms involving derivatives of $\cot\frac{v-u}{2}=\cot\gamma$ should be evaluated appropriately as $\gamma_O\to0$. 
This would eventually require calculating fourth order derivatives of $V$ and, consequently,  require additional explicit data (additional with respect to the lower schemes, e.g., second order derivatives of $V$) on the light cone.
In this paper we do not pursue this higher order scheme further and leave it here with the indication of how it could be completed.

\bibliography{main}{}

\end{document}